\documentclass[MSNbibl,number,dvips]{arxstspdf}
\usepackage{flushend}
\usepackage{stfloats}
\usepackage{graphicx}

\makeatletter
\setattribute{abstract}{width}{405pt}
\setattribute{keyword}{width}{405pt}

\makeatother

\volume{26}
\issue{4}
\pubyear{2011}
\firstpage{543}
\lastpage{563}
\doi{10.1214/11-STS373}

\begin{document}
\begin{frontmatter}
\vspace*{6pt}
\title{In Search of the Black Swan: Analysis of the Statistical Evidence of Electoral Fraud in Venezuela}
\runtitle{Analysis of the Statistical Evidence of Electoral Fraud in Venezuela}

\begin{aug}
\author{\fnms{Ricardo} \snm{Hausmann}}
\and
\author{\fnms{Roberto} \snm{Rigobon}\corref{}\ead[label=e1]{rigobon@mit.edu}}
\runauthor{R. Hausmann and R. Rigobon}

\affiliation{Harvard University and Massachusetts Institute of Technology NBER}

\address{Ricardo Hausmann is the Professor of the Practice of Economic Development and
Director of Center for International Development and the Harvard Kennedy School of Government, Harvard, Mailbox
34,
79 JFK Street,
Cambridge, Massachusetts 02138, USA.
Roberto Rigobon is the Society of Sloan Fellows Professor of Applied Economics at the Sloan School of Management,
Massachusetts Institute of Technology,
Room E62 516,
100 Main Street,
Cambridge, Massachusetts 02139, USA.}

\end{aug}

\begin{abstract}
This study analyzes diverse hypotheses of electronic fraud in the Recall Referendum
celebrated in Venezuela on August 15, 2004. We define fraud as the difference between the
elector's intent, and the official vote tally. Our null hypothesis is that there was no
fraud, and we attempt to search for evidence that will allow us to reject this hypothesis.
We find no evidence that fraud was committed by applying numerical maximums to machines in
some precincts. Equally, we discard any hypothesis that implies altering some machines and
not others, at each electoral precinct, because the variation patterns between machines at
each precinct are normal. However, the statistical evidence is compatible with the
occurrence of fraud that has affected every machine in a single precinct, but
differentially more in some precincts than others. We find that the deviation pattern
between precincts, based on the relationship between the signatures collected to request
the referendum in November 2003 (the so-called, Reafirmazo), and the YES votes on August
15, is positive and significantly correlated with the deviation pattern in the relationship
between exit polls and votes in those same precincts. In other words, those precincts in
which, according to the number of signatures, there are an unusually low number of YES
votes (i.e., votes to impeach the president), is also where, according to the exit polls,
the same thing occurs. Using statistical techniques, we discard the fact that this is due
to spurious errors in the data or to random coefficients in such relationships. We
interpret that it is because both the signatures and the exit polls are imperfect
measurements of the elector's intent but not of the possible fraud, and therefore what
causes its correlation is precisely the presence of fraud. Moreover, we find that the
sample used in the audit conducted on August 18 was neither random nor representative of
the entire universe of precincts. In this sample, the Reafirmazo signatures are associated
with 10 percent more votes than in the non-audited precincts. We built 1,000 random samples
in non-audited precincts and found that this result occurs with a frequency lower than 1
percent. This result is compatible with the hypothesis that the sample for the audit was
chosen only among those precincts whose results had not been altered.
\end{abstract}


\end{frontmatter}

\section{Introduction}

This study presents a statistical evaluation of the results of the August 15, 2004 Recall
Referendum on President Hugo Ch\'{a}vez's mandate. From the morning of August 16, 2004, when
the CNE (Consejo Nacional Electoral) announced the results, opposition spokespersons
expressed doubts about the validity of these results, and argued that an electronic fraud
had been committed. These doubts had not been cleared up with the passing of time.

At the time, S\'{u}mate---a Venezuelan NGO that had organized the collection of
signatures to request the referendum and monitored its execution---reques\-ted
that we do a statistical analysis to verify if the available information is compatible with
the hypothesis of fraud or if, on the contrary, it rejects this hypothesis. S\'{u}mate provided
the data used in this study but gave us complete autonomy over the conduct of our research.

We were informed that the presumption of fraud is based on the following elements:
\begin{longlist}[(1)]
        \item[(1)] The adoption of a new automated voting system in spite of the fact that the opposition had
requested a manual tally.
        \item[(2)] The voting machines left a paper trail by printing ballots that allowed each elector to
verify that the machine had counted his vote adequately. These ballots were collected in
boxes. However, the CNE did not allow the boxes to be opened and counted. Instead, it
performed a so-called ``hot'' audit of 1 percent of the
machines on the evening of the election. Moreover, the CNE decided that the number of boxes
to be opened would be chosen by a random number generator program run on its own computer.
        \item[(3)] After a difficult negotiation, the CNE allowed the Organization of American States and the
Carter Center to participate as observers in every phase of the process except for access
to the central computer server that communicated with each machine in each voting precinct.
No witness from the opposition was granted access to that room either. Only two people were
allowed in that room until the results were ready.
        \item[(4)] The adopted technology allowed---in fact re\-quired---bidirectional communication between
the\break central servers and the voting machines. This bidirectional communication occurred.
This is different from the information that was provided to opposition negotiators about
the nature of the technology involved.
        \item[(5)] Contrary to what was initially stipulated, the voting machines communicated with the
central ser\-ver before printing the results in a document called Acta. This opens the
possibility that the machines were instructed to print a result different from the one
expressed by the voters.
        \item[(6)] On August 15, 2004, different organizations, including S\'{u}mate, conducted exit polls in a
number of precincts. To assure its quality, S\'{u}mate's poll was conducted
with the assistance of the firm Penn, Shoen and Berland. Its results were radically
different from official figures. The same thing occurred with the exit poll conducted by
``Primero Justicia,'' a political party. The database
of both surveys was given to us to conduct this study.
        \item[(7)] The ``hot-audit'' conducted at dawn on August 16, 2004
was not carried out to the satisfaction of either the opposition or the international
observers. Only 78 of the 192 boxes stipulated were counted. The opposition only attended
28 counts, and the international observers were only present in less than 20.
        \item[(8)] As requested by the international observers, a~second audit was conducted on August 18.
The opposition did not participate in this audit because its conditions were not met; for
example, the electoral materials were not delivered to a centralized location before
choosing the boxes to be opened and there was no verification that the boxes selected had
not been tampered with. Instead, the boxes were chosen 24 hours before they were opened,
which in theory would give time for them to be altered. Notably, the CNE did not use the
random number generator program proposed by the Carter Center, and instead insisted on
using its own program run on its own computer and started with a seed defined by a
pro-government member of the CNE. This raises doubts over whether the sample selected was
truly a~random one, or that the sample was unknown before the voting started.
\end{longlist}

All these facts raise the possibility of an electronic fraud in which the machines printed
outcomes different from the real count. This could in theory have been done through
software alterations, or through electronic communications with the computer hub.

Our main findings are the following. First, the paper finds that the sample used for the
audit of August 18, which was observed by the OAS and the Carter Center, was not randomly
chosen. In that sample, the relationship between the votes obtained by the opposition\vadjust{\goodbreak} on
August 15 and the signatures gathered requesting the Referendum in November 2003 was 10
percent higher than in the rest of the boxes. We calculate the probability of this taking
place by pure chance at less than 1 percent. In fact, we create 1,000 samples of
non-audited precincts to prove this.

This result opens the possibility that the fraud was committed only in a subset of the
4,580 automated precincts, say 3,000, and that the audit was successful because it directed
the search to the 1,580 unaltered precincts. This sheds new light on the fact that the
Electoral Council did not accept the use of the random number generator proposed by the
Carter Center and under these conditions one can infer why the Carter Center could not
identify the fraud with the audit they observed. In other words, before the voting process
starts the random seed might be known, and therefore, the computer only changes the
machines that ex-ante knows that have no chance of being audited. The machines audited look
like a random sample regarding regions, social characteristics, etc. except for the fact
that they were not affected by the fraud.

In addition, we develop a statistical technique to identify whether there are signs of
fraud in the data. To do so, we depart from previous work on the subject that was based on
finding patterns in the number of votes per machine or precinct. Instead, we look for two
independent variables that are imperfect correlates of the intention of voters. Fraud is
nothing other than a deviation between the voters' intention and the actual count. Since
each variable used is correlated with the intention, but not with the fraud, we can develop
a test as to whether fraud is present. In other words, each of our two independent measures
of the intention to vote predicts the actual number of votes imperfectly. If there is no
fraud, the errors these two measures generate would not be correlated, as they each would
make mistakes for different reasons. However, if there is fraud, the variables would make
larger mistakes where the fraud was bigger and hence the errors would be positively
correlated. The paper shows these errors to be highly correlated and the probability that
this is pure chance is again less than 1 percent.

The first variable we use is the number of registered voters in each precinct that signed
the recall petition in November, 2003. This clearly shows intent to vote yes in a future
election but it does so imperfectly. Our second measure is the exit poll conducted by Penn,
Schoen and Berland and complemented\vadjust{\goodbreak} with an independent exit poll conducted by Primero
Justicia. This is also an imperfect measure as it depends on potential biases in the
sample, differences in the skill of the interviewer, etc. But this source of error should
not be correlated at the precinct level with the one that affects the signatures.
Therefore, it is very telling that in the precincts where the Penn, Schoen and Berland exit
poll makes bigger mistakes is also where the number of petitioners suggests that the Yes
votes would be higher.

\begin{table*}[t]
\tablewidth=350pt
\caption{Comparison between electoral results and S\'{u}mate's and Primero Justicia's exit polls}
\label{tbl1}
\begin{tabular*}{350pt}{@{\extracolsep{4in minus 4in}}lcc@{}}
\hline
 & \textbf{Unweighted} & \textbf{Weighted} \\
\hline
Percentage of YES votes at the precinct level & 37.0\% & 41.1\% \\
Percentage of YES in S\'{u}mate's exit poll & 59.5\% & 62.0\% \\
Percentage of YES votes where S\'{u}mate did their exit poll & 42.9\% & 45.0\% \\
Percentage of YES in PJ's exit poll & 62.6\% & 61.6\% \\
Percentage of YES votes where PJ did their exit poll & 42.9\% & 42.7\% \\
Percentage of YES in S\'{u}mate${}+{}$PJ exit polls & 61.3\% & 62.2\% \\
Percentage of YES votes where S\'{u}mate${}+{}$PJ did their &  &  \\
\quad exit polls & 43.1\% & 44.2\% \\
\hline
\end{tabular*}
\end{table*}

This evidence is troubling because it resonates with three facts about the conduct of the
election. First of all, contrary to the agreed procedure, the voting machines were ordered
to communicate with the election computer server \textit{before }printing the results.
Second, contrary to what had been stated publicly, the technology utilized to connect the
machines with the computer hub allowed two-way communication and this communication
actually took place. This raises the possibility that the hub could have informed the
machines what numbers to print, instead of the other way around. Finally, after an arduous
negotiation, the Electoral Council allowed the OAS and the Carter Center to observe all
aspects of the election process except for the central computer hub, a place where they
also prohibited the presence of any witnesses from the opposition. At the time, this
appeared to be an insignificant detail. Now it looks much more meaningful.

The structure of the paper is as follows. First, we describe the evidence coming from the
exit polls. We show that the difference between the exit polls and the actual vote is not
likely to have been caused by a sampling error, due for example, to an over-representation
of anti-Chavez precincts, but instead to a generalized but variable difference, precinct by
precinct.

Second, we discuss some of the previous evidence of fraud and its validity. We address the
popular so-called ``topes'' hypothesis. According to
this theory, machines were ordered not to surpass a certain maximum number of Yes votes. If
this was the case, there should be an unusually large number of repeated Yes totals in each
precinct and the repeated number should also be the maximum Yes vote total in the precinct.
We show that the frequency with which the repeated number is also the maximum Yes vote of
the precinct is consistent with a random event---which means that it does not
constitute evidence of fraud in our view. We then move on to study whether the variance of
results at the precinct level is unusual.\vadjust{\goodbreak} This would be the case if some but not all
machines were manipulated at the precinct level. We find the variance at the precinct level
to be if anything smaller than would be expected by pure chance. Again, we do not find
evidence of fraud in this dimension. In the end, the objective of this section is to take a
balanced view to the discovery of fraud.

The next section develops our test for fraud using our two independent but correlated
measures of voters' intent. We then move on to test whether the sample used for the audit
of August 18 was random. The final section concludes.

\section{Discussion on the Earlier Evidence of~Fraud}\label{sec2}

\subsection{Exit Polls Versus Votes: Analysis of the Differences}\label{sec2.1}

The first evidence of potential irregularities in the election count derives from the exit
polls conducted independently by S\'{u}mate and Primero Justicia (PJ). As shown in Table \ref{tbl1},
according to the CNE, 41.1 percent of voters voted YES to impeach the president. On the
other hand, in the S\'{u}mate and PJ surveys, the weighted projections were 62.0 and 61.6
percent, respectively, a difference of more than 20 points.

We check whether this difference is due to the fact that the sample chosen by S\'{u}mate and
Primero Justicia was not representative of the electoral universe. In other words, we check
whether the problem arises because of an over-representation of precincts in favor of the
YES vote in relation to those in favor of the NO. We show that this is not the source of
the problem. As shown in Table~\ref{tbl1}, according to the CNE the percentage obtained by the YES
in the precincts surveyed by S\'{u}mate was 45.0 percent, while in PJ's sample
the result was 42.7 percent. In other words, in the sample chosen by both organizations,
the result reported by them differs from the official tally by more than 17 percentage
points. Hence, the difference in the results is not principally due to the sample
composition but to a systematic difference across the sample where the exit polls were
conducted.

\begin{figure}

\includegraphics{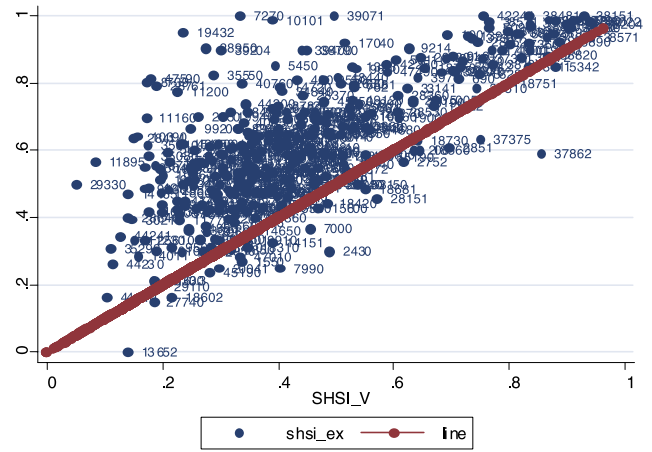}

\caption{Exit polls versus electoral result: percentage of the YES by precinct.}
\label{fig1}
\end{figure}

To illustrate this problem more clearly, in Figure~\ref{fig1} we show the percentage of votes and
the survey results for the 340 precincts surveyed by both groups. If the surveys were
perfect, the points would align in a ray from the origin with a 45 degree slope (drawn in
the graph). In other words, where the YES option received respectively 10 percent, 50
percent or 80 percent, the surveys would show the same result. If the points in the graph
are above the 45 degree line, it means that the poll overestimates the result in that
precinct. If the points are below, the poll underestimates it.

As can be seen, the bulk of the 342 precincts polled are above the 45 degree line.
Moreover, the graph indicates that the differences between the votes and the surveys are
very variable among precincts. The distances to the 45 degree line are largest in places
where the YES option garnered between 20 and 40 percent.

This analysis has the following implications. First, it indicates that the difference
between the surveys and the votes is not due, in any important way, to problems in the
selection of the precincts to be included in the survey. Second, the analysis implies that
the difference may be due to one of the two reasons, or to a combination of both. It may be
due to a~generalized failure in both surveys in each precinct, or to a quite generalized
and nonlinear manipulation of the results. It will be a challenge of the statistical work
to distinguish between these two hypotheses and investigate which is the right
one.\looseness=1

\subsection{The Caps or ``Topes'' Hypothesis}\label{sec2.2}

The fraud hypothesis most discussed in Venezuela has been based on the idea that numerical
caps were imposed on the amount of YES votes that could be allowed in a precinct and that
the overflow of YES votes would be switched into NO votes. In this section we evaluate this
hypothesis.

To analyze the feasibility of this hypothesis we examine how many times the
number of YES and NO votes are repeated at the precinct level in the CNE's
database, which contains 19,062 automated machines (see Table~\ref{tbl2}).

The repetition of the YES count occurs with a~frequency of 9.8 percent while that of the
NO occurs with a~frequency of 7.7 percent. We do not test whether this frequency is
unusually high or low.\footnote{
Jonathan Taylor from Stanford University has argued that it is unusually high. See
\texttt{%
\href{http://www-stat.stanford.edu/\textasciitilde jtaylo/venezuela/}%
    {http://www-stat.stanford.}
\href{http://www-stat.stanford.edu/\textasciitilde jtaylo/venezuela/}%
    {edu/\textasciitilde jtaylo/venezuela/}}.}
However, the relatively
high frequency is at least in part due to the fact that the number of electors as well as
the voting percentage tends to be very similar among machines in the same precinct. The
fact that the repeated YES totals occur with a slightly higher frequency than the NO is at
least in part due to the fact that YES has a lower percentage of votes. Let us illustrate
this point with an example. Suppose the preference for the YES vote in a single precinct is
approximately 40 percent and the number of voters at each machine is 100. A 5 percent
variation would imply 2 votes, so the expected result in each machine could be between 38
and 42. The result could be in some\vadjust{\goodbreak} of the five numbers included in that interval. On the
contrary, the same percent variation for the NO would yield a~variation between 57 and 63
votes, which gives seven possible numbers. Since the amount of possible numbers is higher for
the NO than for the Yes, it is logical the latter would repeat less frequently.

\begin{table}[t]
\tabcolsep=0pt
\caption{Number of YES and NO total votes per machine that are repeated in the same precinct}
\label{tbl2}
\begin{tabular*}{\tablewidth}{@{\extracolsep{4in minus 4in}}lccc@{}}
\hline
\textbf{Variable} & \textbf{Number of machines} & \textbf{Numbers repeated} & \textbf{Frequency} \\
\hline
Si & 19,062 & 1,875 & 9.8 \\
No & 19,062 & 1,472 & 7.7 \\
\hline
\end{tabular*}
\vspace*{-3pt}
\end{table}

\begin{table}[t]
\caption{Maximum and non-maximum numbers repeated per voting tome at the precincts}
\label{tbl3}
\begin{tabular*}{\tablewidth}{@{\extracolsep{4in minus 4in}}lccc@{}}
\hline
\textbf{Machines per precinct} & \textbf{Non-maximum} & \textbf{Maximum} & \textbf{Total} \\
\hline
\phantom{0}2  & \phantom{00}0   & 64 & \phantom{0}64 \\
\phantom{0}3  & \phantom{0}58  & 66 & 124 \\
\phantom{0}4  & 161 & 80 & 241 \\
\phantom{0}5  & 144 & 54 & 198 \\
\phantom{0}6  & 230 & 46 & 276 \\
\phantom{0}7  & 221 & 46 & 267 \\
\phantom{0}8  & 197 & 14 & 211 \\
\phantom{0}9  & 151 & \phantom{0}4  & 155 \\
10 & \phantom{0}97  & \phantom{0}8  & 105 \\
11 & \phantom{0}85  & \phantom{0}2  & \phantom{0}87 \\
12 & \phantom{0}52  & \phantom{0}2  & \phantom{0}54 \\
13 & \phantom{0}36  & \phantom{0}0  & \phantom{0}36 \\
14 & \phantom{0}18  & \phantom{0}0  & \phantom{0}18 \\
15 & \phantom{0}20  & \phantom{0}0  & \phantom{0}20 \\
16 & \phantom{00}7   & \phantom{0}0  & \phantom{00}7 \\
17 & \phantom{00}6   & \phantom{0}0  & \phantom{00}6 \\
18 & \phantom{00}6   & \phantom{0}0  & \phantom{00}6 \\[5pt]
Total & 1,489 & 386 & 1,875 \\
\hline
\end{tabular*}
\vspace*{-3pt}
\end{table}

More importantly, the cap hypothesis implies that the number that repeats itself is also
the maximum from the precinct and that the difference is assigned to the NO. For this, it
is necessary that the repeated number also be the maximum YES vote in the precinct. We
study this hypothesis in Table \ref{tbl3}. The table includes all precincts in which repeated
numbers are observed and classifies them by the number of voting machines in the precinct.
Column one shows the number of machines per precinct. Column 2 shows the number of repeated
numbers that are not the maximum of the precinct, as required by the theory. The third
column shows the number of repeats that are the maximum, while the final column adds the
two.\vadjust{\goodbreak}

If the repeated number was randomly distributed, it would occur with a frequency equal to
1$/$(Number of machines -- 1). For example, in the case of pre\-cincts with two
machines, the repeated number is simultaneously the maximum and the minimum, for there is
only one number. In the case of three machines, the probability that the repeated number is
the maximum is 50 percent.

As we see in Table \ref{tbl3}, 66 is not very far from being half of 124. In the case of five
machines, 54 is not far from being one-fourth of 198. We conclude that if there was fraud, this
was not done through the imposition of numerical caps to the YES votes in the machines of a precinct.

\subsection{Variance Analysis of the Within-Precinct Results}\label{sec2.3}

The caps hypothesis, if true, would also affect the percentage difference in the results
of the machines belonging to the same precinct. This is due to the fact that the amount of
voters per machine varies due to differences in the abstention rate or in the number of
electors assigned to each machine. This variation would show in the number of NO votes, and
therefore would create a source of variation in the results across machines of the same
precinct. This hypothesis and any other hypothesis that is based on the idea of altering
some machine more than others at the precinct level can be tested.

In each precinct, voters are distributed to machines according to the last two digits in
their identity card number (\textit{c\'{e}dula de identidad}). This allows each
machine to be a random sample of the precinct's voters because the last digits in their
identity card are not correlated with any variable relevant to the voting decision. This
limits the possible distance between the results from two machines from the same precinct.
To illustrate this, consider how opinion surveys are done in any country. A random sample
is chosen---usually of a thousand or two thousand people---and the outcome is
used to predict the results of millions of voters. In other words, a representative sample
composed of a miniscule fraction of the electorate is used to predict the outcome of the
whole. In the case of a precinct we are taking a much smaller and homogeneous universe than
a country and we are dividing the population randomly according to the number of machines
in the precinct. For example, in the case of a precinct with five machines, each machine
represents approximately 20 percent of the total population of the precinct. In addition,
in the case of this referendum, the options were limited to two: YES or NO. This imposes a
condition for the standard deviation of the number of votes per machine. Suppose that in a
machine, there were $N$ votes cast and the probability that each vote is a YES is
$p$. Then, the number of YES votes is a binomial random variable with expectation
$Np$ and with standard deviation equal to $\sqrt{p(1 - p)N}$. To illustrate this,
take the case in which $p$ is the probability that
an elector will vote YES in a given precinct, is equal to 50 percent and~$N$ is 400. In this
case, the standard deviation would be 10 votes. The coefficient of variation (or the
standard deviation of the percentage vote) would be 10 divided by 400, meaning, 2.5
percent. Given this, the typical deviation among machines in the same precinct must be
compatible with this rule. If, for example, within a precinct the results of some machines
were changed by 10 percent while the others were left unaltered, then we would see an
increase in the deviation among all machines that would be four times the expected standard
deviation of 2.5 percent. This would be abnormal.

One implication of this result is that the caps or
``topes'' theory would also violate the expected
distribution of a binomial. If numerical caps were assigned to each machine in a precinct,
the variation of the number of voters per machine would affect the number of NO votes and
therefore alter the percentage results in a manner that would increase the dispersion of
the results and cause these to violate the binomial rule.

\begin{figure}

\includegraphics{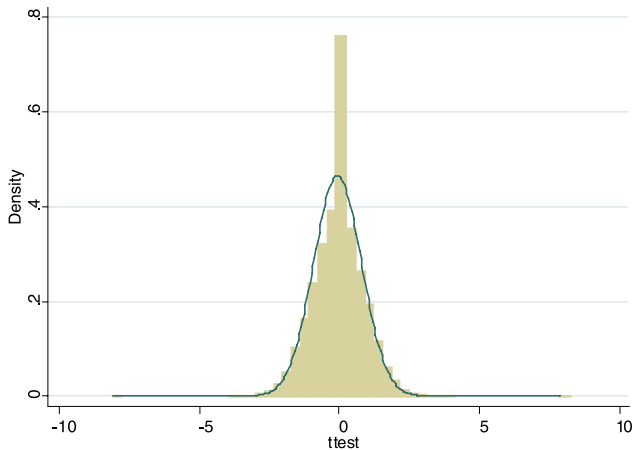}

\caption{Distribution of the deviation of results of machines relative to the precinct mean relative to the predicted standard deviation.}
\label{fig2}
\end{figure}

To verify if the CNE vote data comply with the standard deviation predicted by
probability theory, we calculate each machine's deviation with respect to the average of
its precinct. Moreover, we divide this number by the standard deviation that would
correspond to a precinct with the actual number of voters and machines. Figure \ref{fig2} presents
our results. It shows a histogram of the percent difference among machines of the same
precinct with respect to the standard deviation expected by the binomial distribution. The
curve reflects the expected theoretical distribution. The bars are the frequency calculated
with the actual data. As can be seen, the coincidence is quite substantial. The graph
indicates that only close to 1 percent of the machines have deviations higher than two times
the expected standard deviation. This frequency is consistent with the theoretical
distribution. In fact, if there is anything surprising about the graph, it is that the
deviations of the results are if anything too small, as can be seen by the large
concentration of results near zero variation.\looseness=-1

This result has two possible interpretations. One is that there was no fraud. The other is
that if fraud was committed, it must have been done by changing every machine in the
precinct by a similar percentage. In fact, a fraud of this kind would not be detected with
the analysis done so far for it would not alter the variance results among machines. Any
hypothesis of fraud that does not comply with this condition would violate the restriction
imposed on the deviation of the results by the binomial distribution.

\section{A Statistical Strategy to Detect the Presence of
Fraud}\label{sec3}

To detect if the data are compatible with the presence of fraud we need to develop a model
and fit it to the data. We define fraud as the difference between the voters' intent and
what the electoral system registered about their decision. We will take as our null
hypothesis the assumption that there was no fraud. We will then develop a test to see if
the null hypothesis can be rejected.

The main challenge is that we cannot observe the voters' intent directly. The statistical
strategy we adopt\-ed begins with finding two sets of independent variables that are
correlated to the voters' intent, but not with the fraud. For our purposes, it is not too
important that our variables do not predict the voters' intent perfectly. Even if they do
so imperfectly, it may still give us a chance to reject the hypothesis of no fraud. Notice
that the worse the quality of the data, the harder it will be to\vadjust{\goodbreak} reject the null
hypothesis,
meaning that bad information makes it harder, not easier, to reject the hypothesis of no
fraud.

To illustrate what we do, we start with a simplified presentation of our approach. In
practice, things are a bit more complicated, but explaining the sources of complexity will
be easier after the fundamental intuition is presented.

Let us take two variables that are correlated to the elector's intent: the number of
signatures in favor of holding a recall referendum that were collected during the December
2003 (called Reafirmazo) drive and the proportion of YES responses in the exit polls. We
use $s_i$ and $e_i$ to denote the number of signatures and the number
of YES responses in the exit poll in the $i$th precinct, respectively. Each one of
these variables is an imperfect measure of the voters' intent on August 15, 2004. The
Reafirmazo was a~public vote. Signatures were observed and identities known. The
motivations to sign are different than voting YES in the referendum. For instance, some
people that signed the petition may have changed their opinion in the intervening months.
Others might have decided not to sign because it was not secret, but may have decided to
vote later given its secrecy. Others may not have been registered in November and hence
could not sign, but were registered by August and hence could vote. The lines in the August
election were particularly long and slow and that may have reduced the number of voters,
etc.

Equally, exit polls are an imperfect measure of the voter's intent. Pollsters may have,
consciously or unconsciously, gathered a biased sample. People may have had more or less
willingness to cooperate with the interview, etc. However, these errors are of a quite
different nature from the errors generated by the relationship between signatures and votes
and hence should not be correlated.

\begin{figure}[t]

\includegraphics{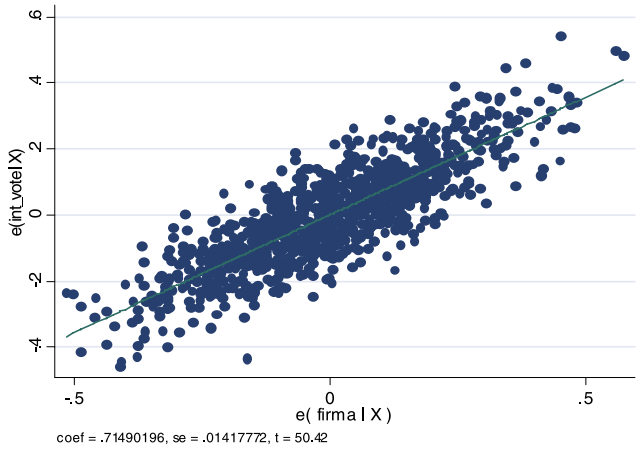}

\caption{Simulated relationship between signatures and voters' intent.}
\label{fig3a}
\end{figure}

\begin{figure}[t]

\includegraphics{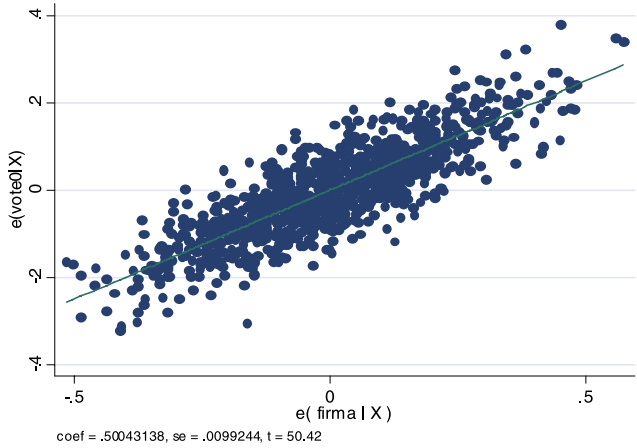}

\caption{Simulated ratio between signatures and votes with fraud proportional to 30 percent of the YES votes.}
\label{fig3b}
\end{figure}

Suppose we have an imperfect measure of the voters' intent in each precinct and we build a
graph relating this variable---say the signatures---and the voters'
intent. As the signatures are an imperfect measure of the voters' intent, the graph will
look like a cloud of dots around some basic relationship (Figure \ref{fig3a}).\footnote{
This graph was built with simulated data using a random number generator. The data were
created supposing that each signature generates 0.7 votes with an error normally
distributed between $+$0.1 and $-$0.1.} Regression analysis can identify the line
that relates the\vadjust{\goodbreak} signature with the voters' intent. The real relationship is 0.7, because
that is how we built the data. The estimated relationship using the simulated data is $0.71 +/-
0.014$, as is indicated by the graph.

We cannot observe the voters' actual intent but can only see the votes registered, and
these, in theory, could be influenced by fraud. Suppose fraud takes place and it is
directly proportional to the numbers of votes in that precinct. For example, let us suppose
that fraud is committed by multiplying the total number of YES votes in a machine by 0.7
and the difference added to the NO votes.

Figure \ref{fig3b} illustrates this case. In this case, the estimated slope is no longer 0.7 but
0.5. In addition, the pattern of errors---that is to say, the distance with
respect to the regression line---looks similar. It reveals no evidence of
fraud. If fraud were committed this way, we would be unable to detect it. In fact, a~fraud
that reduces a fixed percentage of YES votes across all machines would practically be
impossible to detect by purely statistical methods without additional information; that is, it
could only be detected using another source of information such as counting the paper
ballots.

\begin{figure}[t]

\includegraphics{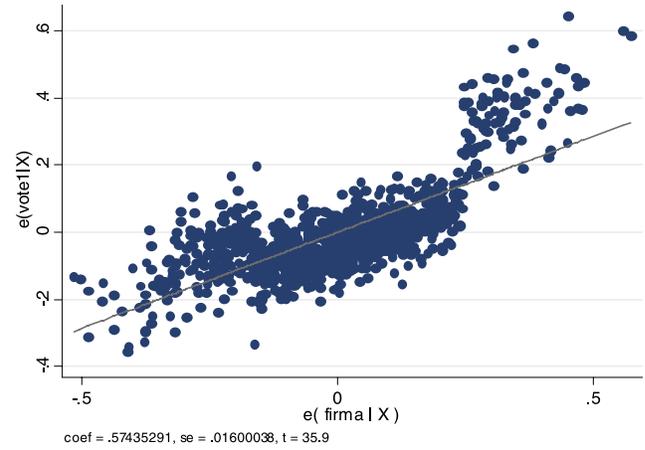}

\caption{Nonproportional fraud.}
\label{fig3c}
\end{figure}

Now, suppose the fraud was not committed in a~proportional manner. For example, suppose it
was committed in some precincts and not in others. Spe\-cifically, suppose fraud consists of
eliminating 30 percent of the YES votes in precincts where signatures were less than 30
percent or more than 70 percent of the registered voters. In this case, the pattern of
errors will have a peculiar shape, as shown in Figure \ref{fig3c}. This peculiarity is not due to
the imperfect nature of the number of signatures as predictor of votes, but is caused by
the presence of fraud.

What happens if we now use a second measure of the voters' intended vote, for example the
exit polls? This is also an imperfect measure of the voters' intended vote and as such when
doing a regression analysis, this will generate errors. Nevertheless, if there is a
nonproportional fraud, this will also generate an irregularity in the errors which will
look similar, that is, will be correlated with the errors in the other relationship. A~positive
and significant correlation would identify nonproportional
fraud.\looseness=1

Note that each measure---the one based on signatures and the one based on exit polls---is
imperfect. Nevertheless, what makes each of them imperfect are factors different and
independent from each other. The exit poll is not influenced by the turnout rate, as people
are interviewed after they vote. The signatures do not depend on the ability or bias of the
interviewer. People could have changed their minds between November and August, but people
do not change their minds for the same reason between the act of voting and the exit
interview. Signing is a~public act and voting is secret, etc. Therefore, errors made by
each measure may be larger or smaller but they should not be correlated. However, if there
is nonproportional fraud, it will influence each of these measures in the same way.
Hence, the errors made by both should be positively correlated. This is the essence of the
method we used.

\section{Instrumental Variable Approach}\label{sec4}

\subsection{Theoretical Considerations}\label{sec4.1}

In this section we derive formally the technique we use. In particular, we show that for a
variety of increasingly complex assumptions about the nature of the fraud the covariance
between the errors of the instrumental variables regression is an appropriate test of the
absence of fraud.\footnote{See Hausman \cite{bib5} and Green \cite{bib4} for an introduction
to instrumental variables. The
original contribution of Instrumental Variables is from Wright \cite{bib8}.}

Assume that the fraud is defined as the difference between the votes for Si actually
collected and an unobservable variable that is the intention of voting of the voters that
showed up. We define the first one as $\nu_i$, the intention of voters as
$\chi_i$, and the fraud as~$\phi_i$:
\begin{equation}\label{eq1}
\nu_i + \chi_i = \phi_i,
\end{equation}
where $i$ indexes precinct. As discussed earlier, two additional measures of the
intention of voters are the number of YES responses in the exit poll in the precinct
($e_i$) and the number of signatures collected in the precinct
($s_i$). These measures, however, are imperfect but can be modeled as
\begin{eqnarray}\label{eq2}
e_i &=& \alpha \chi_i + \varepsilon_i,
\\\label{eq3}
s_i &=& \beta \chi_i + \eta_i,
\end{eqnarray}
where we are assuming that the exit polls are possibly a biased estimate of the intention
to vote: $\alpha$~can be smaller than 1. The signatures ($s_i$'s),
as well, could be a biased measure. Both equations have an error
($\varepsilon_i$ and $\eta_i$) that takes into account the fact
that both the exit polls and the signatures are very imperfect measures of the voter's
intentions---even the biased measured intentions. We assume that these errors
are uncorrelated among themselves and with the fraud.\footnote{This is a reasonable assumption
considering that the signatures were collected at
different times and conditions than the exit polls.}

How can we detect fraud? Fraud can only affect the actual votes, not the exit polls, nor
the signatures. In other words, fraud is a displacement of the distribution of votes that
is not present in the other two measures. Statistically, this means that fraud could be
detected by using the exit polls and the signatures as predictors of the voting process and
analyzing the correlation structure of the residuals. Under the assumption that all
residuals are uncorrelated---which makes sense given the definitions we have
adopted---then the correlation of residuals is an indication of the magnitude
of fraud.

The particular procedure used to detect fraud is the following:
\begin{longlist}[(1)]
        \item[(1)] Estimate the regression of $\nu_i$ on $e_i$ plus controls and
recover the residual. This residual has two components: the fraud and the errors in
variables residual due to the fact that the exit polls are noisy.
        \item[(2)] Estimate the regression of $\nu_i$ on $s_i$ plus controls and
recover the residual. This residual has two components: the fraud and the errors in
variables residual due to the fact that the signatures are an imperfect measure of the
intention of voters.

Notice that these two residuals are correlated: first, because both have fraud as an
unobservable component, and second, because the right-side variables are correlated
and there are errors in variables in the regression.
        \item[(3)] Estimate the regression of $\nu_i$ on $e_i$ plus controls using
$s_i$ as an instrument. Recover the residual. Notice that in our model, because
$\eta_i$ is uncorrelated with $\varepsilon_i$ and
$\phi_i$, we can use $s_i$ as an instrument to correct for the
error in variables.
        \item[(4)] Using the same logic, estimate $\nu_i$ on $s_i$ plus controls,
and using $e_i$ as the instrument. Recover the residual. In this case, because
the two coefficients are supposed to have solved the problem of error in variables the
residuals can only be correlated if there is a common component---which in our
case is the definition of fraud.
\end{longlist}

This procedure actually detects how important fraud is. The next section first explains
why this procedure indeed is able to identify fraud. After that we also analyze
the\vadjust{\goodbreak}
possibility that fraud is correlated with signatures---which is likely given
what we have argued about the stochastic properties of the votes per machine and precinct.
Finally, we present evidence.

\subsection{OLS Estimation (No Correlation Between Fraud and Intention to
Vote)}\label{sec4.2}

Solving for $\chi_i$ in (\ref{eq2}),
\begin{equation}\label{eq4}
\chi_i = \frac{1}{\alpha}e_i - \frac{1}{\alpha}\varepsilon_i,
\end{equation}
substituting in (\ref{eq1}) we find
\begin{equation}\label{eq5}
\nu_i = \theta e_i + \zeta_{1,i},
\end{equation}
where
\[
\zeta_{1,i} = \phi_i - \frac{1}{\alpha}\varepsilon_i.
\]
For this model, the estimate of the slope coefficient, $\theta$, by OLS is
given by
\begin{equation}\label{eq6}
\theta_{\mathrm{ols}} = \frac{\alpha \operatorname{var} (\chi_i)}{\alpha^2 \operatorname{var}(\chi_i) + \operatorname{var}(\varepsilon_i)}
\end{equation}
which is always smaller than $1/\alpha$---the true coefficient. This means that
the residual from the regression ($\zeta_{1,i}$) is
\begin{equation}\label{eq7}
\zeta_{1,i} = \phi_i + \biggl(\frac{1}{\alpha} - \theta_{\mathrm{ols}}\biggr)e_i - \frac{1}{\alpha}\varepsilon_i.
\end{equation}
We can proceed in the same manner but now considering $s_i$ as opposed to
$e_i$. We solve for the intention of voting in equation (\ref{eq3}). The slope
coefficient is
\begin{equation}\label{eq8}
\pi_{\mathrm{ols}} = \frac{\beta\operatorname{var}(\chi_i)}{\beta^2\operatorname{var}({\chi_i})
 + \operatorname{var}(\eta_i)}
\end{equation}
which is always smaller than $\frac{1}{\beta}$.  The residual is given by
\begin{equation}\label{eq9}
\zeta_{2,i} = \phi_i + \biggl(\frac{1}{\beta} - \pi_{\mathrm{ols}}\biggr)s_i - \frac{1}{\beta}\eta_i.
\end{equation}
Notice that the two residuals are correlated. Under the assumption that
$\varepsilon_i$ and $\eta_i$ are uncorrelated, and also
uncorrelated with fraud, there are two components that create the correlation among these
residuals: fraud, and the errors-in-variable bias:
\begin{eqnarray*}
\operatorname{cov}(\zeta_{1,i},\zeta_{2,i}) &=& \operatorname{var}(\phi_i)
\\
&&{} +
        \biggl(\frac{1}{\alpha} - \theta_{\mathrm{ols}}\biggr)\biggl(\frac{1}{\beta} - \pi_{\mathrm{ols}}\biggr)\operatorname{cov}(e_i,s_i).
\end{eqnarray*}
The first term is the variance coming from fraud, while the second term comes from the
variance due to the error in variables that is present in both $e_i$ and
$s_i$. Notice that we are assuming that the errors in variables are independent.
The covariance arises because the error in variables downward biases both coefficients
($\theta_{\mathrm{ols}} < \frac{1}{\alpha}$ and $\pi_{\mathrm{ols}} < \frac{1}{\beta}$) and because the exit polls and the
signatures are correlated through the voter's intention.

\subsection{Instrumental Variables}

Under our assumptions, we have an easy solution to the error in variables in both
regressions. Notice that $\eta_i$ and $\varepsilon_i$ are
uncorrelated and that $\eta_i$ is uncorrelated with
$\phi_i$. Additionally, $e_i$ and $s_i$ are
correlated because both measure the same factor ($\chi_i$). This means that
$s_i$ can be used for instrumenting~$e_i$ and $e_i$ for
instrumenting $s_i$. The outcome is as follows:
\begin{equation}\label{eq10}
\nu_i = \theta e_i + \phi_i - \varepsilon_i.
\end{equation}
The IV estimate is
\begin{eqnarray*}
\theta_{\mathrm{iv}} &=& \frac{\operatorname{cov}(s_i v_i)}{\operatorname{cov}(s_i e_i)}, \\
\theta_{\mathrm{iv}} &=& \frac{\beta\operatorname{var}(\chi_i)}{\alpha\beta\operatorname{var}(\chi_i)}, \\
\theta_{\mathrm{iv}} &=& \frac{1}{\alpha},
\end{eqnarray*}
which means that the residual is
\begin{equation}\label{eq11}
\zeta_{1,i} = \phi_i - \frac{1}{\alpha}\varepsilon_i.
\end{equation}
Notice that now the errors-in-variables component has disappeared. Similarly, if we run the
regression for votes on signatures and using the exit polls as instrument, we find:
\begin{equation}\label{eq12}
\nu_i = \pi s_i + \phi_i - \frac{1}{\beta}\eta_i.
\end{equation}
The IV estimate is
\begin{eqnarray*}
\pi_{\mathrm{iv}} &=& \frac{\operatorname{cov}(e_i v_i)}{\operatorname{cov}(e_i e_i)}, \\
\pi_{\mathrm{iv}} &=& \frac{\operatorname{var}(\chi_i)}{\beta\operatorname{var}(\chi_i)}, \\
\pi_{\mathrm{iv}} &=& \frac{1}{\beta},
\end{eqnarray*}
which means that the residual is
\begin{equation}\label{eq13}
\zeta_{2,i} = \phi_i - \frac{1}{\beta}\eta_i.
\end{equation}
The correlation between the residuals of the two IV regression is now
\begin{equation}\label{eq14}
\operatorname{cov}(\zeta_{1,i}, \zeta_{2,i}) = \operatorname{var}(\phi_i).
\end{equation}
So, a simple test is to compare these two covariances, and determine if they are
statistically different from zero. Furthermore, if the covariance of the IV residuals is
different from zero, then we have an estimate of the importance of the fraud.\footnote{
This procedure is in the spirit of the recent literature on identification through
heteroskedasticity (Rigobon \cite{bib6}). The classical reference on identification is Fisher
\cite{bib3}.}

\subsection{Estimation When There is Correlation Between Fraud and Intention to Vote}

The previous exercise has assumed that fraud is uncorrelated with the signatures as a
measure of the intent to vote, but as we have argued in the previous section, this is
unlikely. In fact, most probably, fraud is correlated with the signatures because the
government used that information in the design of the fraud. Let us repeat the previous
exercise allowing for any covariance structure between fraud and the signatures. We assume
that the number of votes in each precinct is $\nu_i = \chi_i + \phi_i + \rho s_i$, where
$\rho$ is the coefficient capturing the correlation between the fraud and the
signatures.

The residual from running OLS of votes on exit polls is
\begin{equation}\label{eq15}
\zeta_{1,i} = \phi_i + \biggl( \frac{1}{\alpha} - \theta_{\mathrm{ols}} \biggr)e_i
- \frac{1}{\alpha}\varepsilon_i + \rho s_i,
\end{equation}
while the residual of running votes on signatures is
\begin{equation}\label{eq16}
\zeta_{2,i} = \phi_i + \biggl( \frac{1}{\beta} - \pi_{\mathrm{ols}} + \rho \biggr)s_i - \frac{1}{\beta}\eta_i.
\end{equation}
Equations~(\ref{eq15}) and (\ref{eq16}) are equivalent to equa-\break tions~(\ref{eq7}) and (\ref{eq9}). Notice that the two
residuals are correlated as before, but now there are two additional terms:
\begin{eqnarray*}
&&\operatorname{cov}(\zeta_{1,i},\zeta_{2,i})
\\
 &&\quad= \operatorname{var}(\phi_i) + \biggl( \frac{1}{\alpha} - \theta_{\mathrm{ols}} \biggr)
 \biggl( \frac{1}{\beta} - \pi_{\mathrm{ols}} \biggr)\operatorname{cov}(e_i, s_i)
  \\
&&\qquad {}+ \biggl( \frac{1}{\alpha} - \theta_{\mathrm{ols}} \biggr) \rho  \operatorname{cov}(e_i, s_i) + \rho^2\operatorname{var}(s_i).
\end{eqnarray*}
That arises from the correlation between fraud and the signatures.

On the other hand, the residual of the IV regression when votes\vadjust{\goodbreak} are projected on the exit
poll and signatures are used as instruments is
\begin{equation}\label{eq17}
\zeta_{1,i} = \phi_i - \frac{1}{\alpha}\varepsilon_i + \rho s_i + \biggl(\frac{1}{\alpha} - \theta_{\mathrm{iv}}\biggr)e_i,
\end{equation}
while the residual of projecting votes on signatures and using the exit polls as
instruments is
\begin{equation}\label{eq18}
\zeta_{2,i} = \phi_i - \frac{1}{\beta}\eta_i + \rho s_i.
\end{equation}
One point worth noticing is that when exit polls are used as instruments, the results are
identical to those in the previous subsection. In other words, whether fraud is correlated or not
with the signatures, makes no difference in the validity and quality of exit polls as
instruments. The fraud is not correlated with the exit polls or their innovations.
Signatures, on the other hand, are correlated with fraud. This makes exit polls a good
instrument for signatures, but signatures are not a good instrument for exit polls.

Let us compare the two covariances: the covariance for the OLS residuals with the
covariance for the IV residuals. The OLS residuals have covariance equal to
\begin{eqnarray}\label{eq19}
\quad &&\operatorname{cov}(\zeta_{1,i},\zeta_{2,i})^{\mathrm{OLS}}\nonumber
\\
&&\quad= \operatorname{var}(\phi_i) + \biggl( \frac{1}{\alpha} -
        \theta_{\mathrm{ols}} \biggr) \biggl( \frac{1}{\beta} - \pi_{\mathrm{ols}} \biggr)\operatorname{cov}(e_i, s_i)
\\
&&{}\qquad + \biggl( \frac{1}{\alpha} - \theta_{\mathrm{ols}} \biggr) \rho  \operatorname{cov}(e_i, s_i) + \rho^2\operatorname{var}(s_i)\nonumber
\end{eqnarray}
while the covariance for the IV estimates is
\begin{eqnarray}\label{eq20}
&&\operatorname{cov}(\zeta_{1,i},\zeta_{2,i})^{\mathrm{IV}}\nonumber
\\
&&\quad= \operatorname{var}(\phi_i) + \rho\biggl(\frac{1}{\alpha} - \theta_{\mathrm{iv}}\biggr)\operatorname{cov}(e_i,s_i)
\\
&&\qquad{}+ \rho^2\operatorname{var}(s_i).\nonumber
\end{eqnarray}
First, notice that as before, if there is no fraud the covariance of the IV residuals
should be zero. Furthermore, the covariance in equation (\ref{eq2}) reflects the different forms of
fraud. The first element ($\operatorname{var}(\phi_i)$) is the component when fraud is
random, while the last two terms capture fraud when it is correlated with the signatures.

Second, the difference between the two covariances is
\begin{eqnarray}\label{eq21}
&&\operatorname{cov}^{\mathrm{OLS}} - \operatorname{cov}^{\mathrm{IV}}\nonumber
\\
&&\quad= \biggl(\frac{1}{\alpha} - \theta_{\mathrm{ols}}\biggr)\biggl(\frac{1}{\beta} - \pi_{\mathrm{ols}}\biggr)\operatorname{cov}(e_i, s_i)
        \\
        &&\qquad{}+ (\theta_{\mathrm{iv}} - \theta_{\mathrm{ols}})^* \rho  \operatorname{cov}(e_i,
        s_i),\nonumber
\end{eqnarray}
where the two terms are easily signed. We know that the error in variables, together with a
negative~$\rho$, implies that both OLS estimates are downward biased. We also
know that a reasonable set of assumptions imply that signatures and exit polls are
positively correlated. Hence, the first term is a multiplication of three positive
elements. Additionally, we know that $\theta_{\mathrm{iv}}$ is closer to
$\frac{1}{\alpha}$ than $\theta_{\mathrm{ols}}$. This means that the term in
brackets is negative, and we have been analyzing only the case in which $\rho$
is negative. Hence, the covariance of the OLS residuals has to be larger than the
covariance of the IV residuals. Notice that if $\rho$ were positive we could not
have made this claim. And there would be circumstances in which the covariance actually
goes up after instrumenting.\looseness=1

\subsection{Results}\label{sec4.2}

In this section we estimate the covariance of the residuals from the OLS and instrumental
variables. Our results are summarized as follows: first, we observe a reduction in the
covariance, as would be consistent with the presence of $\rho$ being negative
(i.e., against the YES vote). Second, we find that $\operatorname{cov}^{\mathrm{IV}}$ is positive and statistically
significant, which is informative of the fact that $\rho$ is significantly
different from zero and there is fraud.

\begin{table*}
\tablewidth=400pt
\caption{Estimate of the equation between votes and signatures, new voters and voters participating}
\label{tbl4}
\begin{tabular*}{230pt}{@{\extracolsep{4in minus 4in}}lccc@{}}
\hline
\textbf{Source} & \textbf{SS} & \textbf{df} & \textbf{MS} \\
\hline
Model & 185.800888\phantom{00} & \phantom{00}3 & 61.9336295\phantom{00} \\
Residual & \phantom{00}5.84296339 & 338 & \phantom{0}0.017286874 \\[5pt]
Total & 191.643852\phantom{00} & 341 & 0.56200543\\
\hline
\end{tabular*}
\begin{eqnarray*}
\mathrm{Number\ of\ obs} &=&     342 \\[-2pt]
F(3, 338) &=&  3582.70  \\[-2pt]
\mathrm{Prob} > F      &=&  0.0000 \\[-2pt]
R\mbox{-squared}     &=&  0.9695 \\[-2pt]
\mathrm{Adj}\ R\mbox{-squared} &=&  0.9692 \\[-2pt]
\mathrm{Root\ MSE}      &=&  0.13148
\end{eqnarray*}
\begin{tabular*}{330pt}{@{\extracolsep{4in minus 4in}}lcccccc@{}}
\hline
$\bolds{\nu_i}$ & \textbf{Coef.} & \textbf{Std. err.} & $\bolds{t}$ & $\bolds{P > |t|}$ & \multicolumn{2}{c}{\textbf{[95\% Conf. interval]}} \\
\hline
$e_i$   &                    0.9942821 &  0.0099034  & 100.40 &  0.000 &     0.974802\phantom{0} &   1.013762\phantom{0} \\
$\mathrm{newvote}_i$ & 0.4604462   &   0.0375\phantom{000}  &  \phantom{0}12.28 &  0.000  &   0.3866834  &  0.5342089 \\
$\mathrm{turnout}_i$ & 0.3311808 &  0.0813913  &   \phantom{00}4.07 &  0.000  &   0.1710835  &  0.4912781 \\
\_$\mathrm{cons}$ & 0.3059669 &  0.0782436  &   \phantom{00}3.91 &  0.000  &   0.1520611  &  0.4598727 \\
\hline
\end{tabular*}
\end{table*}

\begin{table*}[b]
\tablewidth=360pt
\caption{Estimate of the relationship between votes and the exit polls}
\label{tbl5}
\begin{tabular*}{230pt}{@{\extracolsep{4in minus 4in}}lccc@{}}
\hline
\textbf{Source} & \textbf{SS} & \textbf{df} & \textbf{MS} \\
\hline
Model & 157.862978\phantom{0} & \phantom{00}3 & 52.6209927\phantom{00} \\
Residual & \phantom{0}33.7808737 & 338 & \phantom{0}0.099943413 \\[5pt]
Total & 191.643852\phantom{0} & 341 & 0.56200543\\
\hline
\end{tabular*}
\begin{eqnarray*}
\mathrm{Number\ of\ obs} &=&     342 \\[-2pt]
F(3, 338) &=&  526.51 \\[-2pt]
\mathrm{Prob} > F      &=&  0.0000 \\[-2pt]
R\mbox{-squared}     &=&  0.8237 \\[-2pt]
\mathrm{Adj}\ R\mbox{-squared} &=&  0.8222 \\[-2pt]
\mathrm{Root\ MSE}      &=&  0.31614
\end{eqnarray*}
\begin{tabular*}{360pt}{@{\extracolsep{4in minus 4in}}lcccccc@{}}
\hline
$\bolds{\nu_i}$ & \textbf{Coef.} & \textbf{Std. err.} & $\bolds{t}$ & $\bolds{P > |t|}$ & \multicolumn{2}{c}{\textbf{[95\% Conf. interval]}} \\
\hline
$e_i$   &   \phantom{$-$}0.9701892 &   0.025357\phantom{0} &    \phantom{$-$}38.26 &  0.000   &  \phantom{$-$}0.9203118 &   \phantom{$-$}1.020067\phantom{0} \\
$\mathrm{newvote}_i$ &  $-$0.6612884 &  0.0868377 &   \phantom{0}$-$7.62 &  0.000 &   $-$0.8320987 &   $-$0.490478\phantom{0}  \\
$\mathrm{turnout}_i$ &   \phantom{$-$}0.4244489 &  0.1957766   &  \phantom{$-$0}2.17 &  0.031  &   \phantom{$-$}0.0393549  &  \phantom{$-$}0.8095429  \\
\_$c_0$ &   \phantom{$-$}0.0722736 &  0.2086177 &     \phantom{$-$0}0.35 &  0.729 &   $-$0.3380789 &   \phantom{$-$}0.4826261 \\
\hline
\end{tabular*}
\end{table*}

It is important to remember that in this procedure we are allowing the exit polls and the
signatures to be imperfect measures of the actual votes. Not only do we allow them to be
noisy, but we also allow them to be biased. So, our results will NOT depend on the fact
that the mean of the exit polls is different from the mean of the votes.

For the estimation, we included other explanatory variables in our analysis. These are:
(i) the number of new voters in each precinct from the time the signatures were collected; and
(ii) the turnout rate in each precinct.

The new voters were unable to take part in the Reafirmazo as they had not been previously
registered to vote. The more new voters there are, the greater the number of votes there
should be. Now then, the percentage of YES votes could increase or diminish according to
the difference in political preferences of the new voters with respect to those registered
previously. As in the previous case, the turnout rate obviously increases the number of
votes and is able to do so in a differentiated manner between the YES and NO options.

The estimated equation is (all variables in  logs):
\begin{eqnarray}\label{eq22}
\nu_i &=& c_0 + c_1 s_i + c_2 \operatorname{newvote}_i\nonumber
\\[-8pt]\\[-8pt]
&&{} + c_3      \operatorname{turnout}_i +
\varepsilon_i,\nonumber
\end{eqnarray}
where $\nu_i$ is the logarithm of the number of YES votes; $s_i$ is
the logarithm of the number of signatures in each precinct; $\mathrm{newvote}_i$ is the
percentage of new voters; $\mathrm{turnout}_i$ is the percentage of voters participating;
and $c_0$, $c_1$, $c_2$ and $c_3$ are
parameters to be estimated. Table~\ref{tbl4} shows the results of our estimates for the 342
(voting) precincts for which we also have exit polls, using the most conventional method:
the squared minimums.

The estimate allows us to explain 97 percent of the variation in votes among (voting)
precincts.  It estimates parameters $ c_0$, $ c_1$, $ c_2$ and $ c_3$ with great precision.  Specifically, $ c_0$ is
the constant, estimated at 0.306.  Parameter $ c_1$ is the elasticity between signatures and
votes and is estimated at almost 1 (in reality it is 0.994).  This implies that if a
precinct has twice as many signatures as another, it obtains on average twice as many
votes.  Parameter $ c_2$ is the elasticity of the YES votes to variations in the percentage of
new voters.  It is estimated at 0.46, which means that if the number of voters in a
precinct increases by 100 percent, the YES votes would increase by 46 percent.  Parameter $ c_3$
is the elasticity of the number of YES votes compared to a change in the voters
participating and is estimated at 0.306, which indicates that a 10 percent increase in the
rate of voters participating would cause a 3.06 percent increase in the number of YES
votes.\looseness=1

This equation does not indicate the actual ratio between the voters' intended vote and its
explanatory variables, but between the latter and the votes published. As in Figure \ref{fig3b},
the possible presence of fraud influences the estimated coefficients, biasing the slopes
downward, and in part is found in the error term.

The second equation we estimated was the ratio between votes and exit polls also for the
342 precincts for which we have data. The equation we estimated is similar to equation (22)
but now using the exit polls ($e_i$) as the proxy for voters' intentions:
\begin{eqnarray}\label{eq23}
\nu_i &=& c_0 + c_1 e_i + c_2 \operatorname{newvote}_i\nonumber
\\[-8pt]\\[-8pt]
&&{}+ c_3      \operatorname{turnout}_i +
\varepsilon_i,\nonumber
\end{eqnarray}
where $e_i$ is the number of YES votes which the poll for this precinct
predicts given the number of voters that actually showed up to vote and the percentage of
YES votes documented in the poll.  The results appear in Table \ref{tbl5}.

\begin{table}
\caption{Analysis of the relationship between the errors in the equations using minimum squares}
\label{tbl6}
\begin{tabular*}{\tablewidth}{@{\extracolsep{4in minus 4in}}ll@{}}
\hline
{Covariance} & ${9.3 \times 10^{-3}}$ \\
Covariance typical deviation & $2.8 \times 10^{-3}$ \\
T-Student on the covariance & 4.1 \\
Probability different from zero & 0.999 \\
Correlation & 0.24 \\
\hline
\end{tabular*}
\end{table}

\begin{table*}[b]
\tablewidth=370pt
\caption{Regression between votes and signatures using exit polls as an instrumental variable}
\label{tbl7}
\begin{tabular*}{240pt}{@{\extracolsep{4in minus 4in}}lccc@{}}
\multicolumn{4}{c@{}}{Instrumental variables (2SLS) regression}\\
\hline
\textbf{Source} & \textbf{SS} & \textbf{df} & \textbf{MS} \\
\hline
Model & 185.741458\phantom{00} & \phantom{00}3 & 61.9138192\phantom{00} \\
Residual & \phantom{00}5.90239422 & 338 & \phantom{0}0.017462705 \\[5pt]
Total & 191.643852\phantom{00} & 341 & 0.56200543\\
\hline
\end{tabular*}
\begin{eqnarray*}
\mathrm{Number\ of\ obs} &=&     342 \\[-2pt]
F(3, 338) &=&  3013.34 \\[-2pt]
\mathrm{Prob} > F      &=&  0.0000 \\[-2pt]
R\mbox{-squared}     &=&  0.9692 \\[-2pt]
\mathrm{Adj}\ R\mbox{-squared} &=&  0.9689 \\[-2pt]
\mathrm{Root\ MSE}      &=&  0.13215
\end{eqnarray*}
\begin{tabular*}{370pt}{@{\extracolsep{4in minus 4in}}lccccccc@{}}
\hline
$\bolds{\nu_i}$ & \textbf{Coef.} & \textbf{Std. err.} & $\bolds{t}$ & $\bolds{P > |t|}$ & \multicolumn{2}{c}{\textbf{[95\% Conf. interval]}} \\
\hline
$s_i$   &   \phantom{$-$}0.9701892 &   0.025357\phantom{0} &   \phantom{$-$}38.26 &  0.000   &  \phantom{$-$}0.9203118 &   \phantom{$-$}1.020067\phantom{0} \\
$\mathrm{newvote}_i$ &  $-$0.6612884 &  0.0868377 &   \phantom{0}$-$7.62 &  0.000 &   $-$0.8320987 &   $-$0.490478\phantom{0}  \\
$\mathrm{turnout}_i$ &   \phantom{$-$}0.4244489 &  0.1957766   &  \phantom{$-$0}2.17 &  0.031  &   \phantom{$-$}0.0393549  &  \phantom{$-$}0.8095429  \\
$c_0$   &  \phantom{$-$}0.0722736 &  0.2086177 &    \phantom{$-$0}0.35 &  0.729 &   $-$0.3380789 &   \phantom{$-$}0.4826261
\\[5pt]
Instrumented: & \multicolumn{6}{@{}l}{$s_i$}         \\
Instruments: &   \multicolumn{6}{@{}l}{$\mathrm{newvote}_i \ \mathrm{turnout}_i \ e_i$}   \\
\hline
\end{tabular*}
\end{table*}

Again, the equation explains a large part of the variance of the logarithm\vadjust{\goodbreak} of votes
(82\%). The estimated elasticity of the voting intentions according to the polls is 0.97.
These estimates could also be biased downward by the presence of fraud, but we are mostly
interested in understanding what the implications are regarding the correlation of the
residuals.

In Table~\ref{tbl6} we present the covariances, their differences and their significance. The
correlation is 24\%, which is surprisingly high. This does not permit us to reject the
fraud hypothesis. In other words, in places where the signatures are proportionally wrong
in the sense of predicting more YES votes than those obtained, the exit polls also
overestimate the YES votes. Since both measurements are independent, the implication is
that what they have in common is fraud.

This is the first result consistent with the fraud hypothesis. Formally, we can say that
we cannot reject the hypothesis that fraud was committed. The presence of this correlation
indicates that there is something in common between the errors committed by the exit poll
and the errors committed by the signatures and this is consistent with a difference between
the elector's voting intent and the official tally.

However, it is possible to argue that the observed correlation might be generated by two
sources.  One is the fact that our measurements of the voter's intent are very noisy or
imperfect and that the errors in such variables might generate problems. The second is that
we suppose fixed coefficients between signatures and votes or between exit polls and votes,
and that these coefficients might be random. This opens the possibility that the
correlation we are finding may have been generated by other factors and not by fraud.

To discard this possibility we estimate using the IV strategy. Table \ref{tbl7} shows the same
equation as Table \ref{tbl4}, but this time it uses the instrumental variables method where the exit
polls is the instrument.

\begin{table*}[t]
\tablewidth=370pt
\caption{Regression between the votes and exit polls using signatures as an instrumental variable}
\label{tbl8}
\begin{tabular*}{250pt}{@{\extracolsep{4in minus 4in}}lccc@{}}
\multicolumn{4}{c@{}}{Instrumental variables (2SLS) regression}\\
\hline
\textbf{Source} & \textbf{SS} & \textbf{df} & \textbf{MS} \\
\hline
Model & 151.228444\phantom{0} & \phantom{00}3 & 50.4094815\phantom{00} \\
Residual & \phantom{0}40.4154074 & 338 & \phantom{0}0.119572211 \\[5pt]
Total & 191.643852\phantom{0} & 341 & 0.56200543\\
\hline
\end{tabular*}
\begin{eqnarray*}
\mathrm{Number\ of\ obs} &=&     342 \\[-2pt]
F(3, 338) &=&  517.96  \\[-2pt]
\mathrm{Prob} > F      &=&  0.0000 \\[-2pt]
R\mbox{-squared}     &=&  0.7891 \\[-2pt]
\mathrm{Adj}\ R\mbox{-squared} &=&  0.7872 \\[-2pt]
\mathrm{Root\ MSE}      &=&  0.34579
\end{eqnarray*}
\begin{tabular*}{370pt}{@{\extracolsep{4in minus 4in}}lcccccc@{}}
\hline
$\bolds{\nu_i}$ & \textbf{Coef.} & \textbf{Std. err.} & $\bolds{t}$ & $\bolds{P > |t|}$ & \multicolumn{2}{c}{\textbf{[95\% Conf. interval]}} \\
\hline
$e_i$                     &    \phantom{$-$}1.176787\phantom{0}  &  0.030827\phantom{0}  &  \phantom{$-$}38.17 &  0.000   &   \phantom{$-$}1.11615\phantom{00} &   \phantom{$-$}1.23742\phantom{00} \\
$\mathrm{newvote}_i$  &   $-$0.6829967  & 0.0949936  &  \phantom{0}$-$7.19 &  0.000  &  $-$0.8698498 &   $-$0.4961437 \\
$\mathrm{turnout}_i$   &     \phantom{$-$}0.1627794  & 0.2148175 &    \phantom{$-$0}0.76  & 0.449  &  $-$0.2597683  &  \phantom{$-$}0.5853271 \\
$c_0$                     &  $-$1.523351\phantom{0}   & 0.250735\phantom{0} &   \phantom{0}$-$6.08 &  0.000  &  $-$2.016549\phantom{0} &  $-$1.03015\phantom{00}
\\[5pt]
Instrumented: & \multicolumn{6}{@{}l}{$e_i$}         \\
Instruments: &   \multicolumn{6}{@{}l}{$\mathrm{newvote}_i \ \mathrm{turnout}_i \ s_i$}   \\
\hline
\end{tabular*}
\end{table*}

Note that the coefficient of the signatures now increases: from 0.994 in the estimate in
Table \ref{tbl4} to 1,013 in Table \ref{tbl7}. This is normal, as the existence of errors or noise in the
data tends to lower the coefficients estimated with the method of Table \ref{tbl4}. On cleaning or
lowering the problem of errors in the data, higher coefficients are usually obtained.

Table \ref{tbl8} re-estimates the same equation as in Table~\ref{tbl5} but using instrumental variables.
This time, the coefficient of the exit poll (\textit{e$_{i}$}) increases from 0.97 to
1.17. This is to be expected as the data of the exit polls, given their nature, are
noisier than the signature data, which is why the method in Table~\ref{tbl5} skews the coefficient
more than in the case of the signatures.

On studying the relationship between the errors in these two equations, we obtain the data
presented in Table \ref{tbl9}. The~analysis shows that even after~using the method of instrumental
variables to correct for problems of errors in variables, the correlation~between errors
generated using signatures and those generated using exit polls diminishes only from 0.24
to 0.17 and remains significantly different from zero.

\begin{table*}[t]
\tablewidth=280pt
\caption{Analysis of the relationship between errors in the two equations used to estimate the number of votes: minimum squares versus instrumental variables}
\label{tbl9}
\begin{tabular*}{300pt}{@{\extracolsep{4in minus 4in}}lcc@{}}
\hline
\textbf{Item} & \textbf{Minimum squares}  &  \textbf{Instrumental variables} \\
 & \textbf{method}  &  \textbf{method} \\
  \hline
Covariance & $9.3 \times 10^{-3}$ & $7.7 \times 10^{-3}$ \\
Typical distortion & $2.8 \times 10^{-3}$ & $2.5 \times 10^{-3}$ \\
Probability different from zero & 0.999\phantom{0000.} & 0.991\phantom{0000.} \\
Correlation & 0.24\phantom{00000.} & 0.17\phantom{00000.} \\
T-Student of covariance & 4.1\phantom{000000.} & 3.1\phantom{000000.} \\
 \hline
\end{tabular*}
\end{table*}

Table \ref{tbl9} summarizes our two main results: First, the correlation of the residuals in the
OLS and in the IV strategies is statistically different from zero. Second, the OLS
covariance and correlation are larger than in the IV. This means that we reject the
hypothesis that there was no fraud, and we reject the hypothesis that fraud was random.

Our strategy has consisted in utilizing two sources of information related to the voters'
intended vote but not to the possible fraud. If we use these\vadjust{\goodbreak} sources or variables
in~estimating the votes imperfectly, then the residual will contain not only the
imperfections of our sources~but also a component associated with fraud. Our
interpretation is that since the imperfections are independent~from each other and the
residuals are correlated, there must be a common factor tying them together, that is, fraud.

\subsection{The Audit}\label{sec4.6}

Any hypothesis of fraud requires an explanation of why the audits that took place did not
find any foul play. While the first audit carried out in the wee hours of the morning of
August 16 failed, the audit conducted on August 18, if it was well carried out, should have
settled the issue. The audit was based~on~opening 150 randomly selected ballot boxes, which
contain the original paper ballots checked by the voters and which thus reflect their real
intended vote. If these boxes were not tampered with~and if they really are a random
sampling~of the universe of precincts, the audit should rule out any presumption of fraud.
So, how could fraud have taken place,~if~the audit did not find it? It should be pointed
out that any hypothesis of fraud which involves changing hundreds of ballot boxes would
constitute a conspiracy involving a large number of\vadjust{\goodbreak} participants and hence would be more
likely to be revealed through disloyalty.

One hypothesis is that fraud was not committed in all precincts but only in a fraction of
them. To give an example, suppose that out of the 4,580 automated precincts used in the
election, 3,000 precincts were altered but~the rest were not. Let us further suppose that
the unaltered 1,580 precincts were\break picked at random. This implies that they would represent
a balanced sample of the country from a regional and social point of view. The same~would
be true of the 3,000~precincts~in which the results~supposedly had been altered. One reason
to do things this way is that it was known (beforehand) that ex-post audits~would be
carried out and that a number of precincts would be checked. To accommodate this, they
would have to be unaffected by fraud.

Note that if fraud is committed in some precincts and not in others, then it will not be
perfectly proportional and the method used in the previous section would detect it. If the selection of the
precincts left unaffected was done in this way, this creates an
important complication but also opens up a great opportunity. The complication is that the
selection of the boxes to be audited could not really be random ex-post. It is critical
that the selection be made among the 1,580 untampered precincts and not\break among the 3,000
tampered ones.~This is only possible if one has control over the random number generator
that selects the boxes to be audited. In this sense, it has to be pointed out that the
National Electoral Council refused to~make use of the random numbers-generating program
proposed by the Carter Center and insisted on the use of their own program installed in
their own computer---which was the one counting the votes and connecting to the
electronic machines.

The opportunity generated by this form of addressing the audit problem is that any sample\vadjust{\goodbreak}
taken of the 1,580 untampered precincts is a representative sample of the country in the
social and regional sense. This~makes it more difficult to know if the sample taken was
really random, as it resembles the country in all the dimensions usually associated~with
representativeness, such as regional or social.

To solve this problem~we must develop a methodology~that allows us to test if the sample
taken for the audit on August 18th really is a random sample. To understand the problem
more clearly, let us call the tampered precincts ``fat'' and the untampered ones ``thin.'' The
sample taken for the audit must be a sample of only ``thin'' precincts,~while the rest of the
precincts are a mixture of ``fat'' and ``thin.'' If we could~``weigh'' the audited precincts, we
would~be able to see that on average the un-audited precincts are ``fatter.'' The problem is
that we need to develop a methodology that can test whether the audited precincts weigh as
much as the others do or whether on the contrary they have a statistically different body
frame.

The method we suggest is as follows. There exists a theorem in statistics that states that
if a ratio applies to an entire unit, any random sampling of the same must have the exhibit
ratio. If~we estimate the ratio of the universe of un-audited precincts and estimate
another for the audited ones, the second cannot be statistically different from the first.
Otherwise, it would not~be a random and representative sample.

To implement this strategy~we again made use of our model that correlates signatures,
voter participation rates and new voters with the number of~actual votes cast. In this
case, because we are not going to use the exit polls in the analysis we can use the data
from all precincts. We estimated this ratio based on the universe of 4,580 precincts and we
looked at the obtained coefficients. We then estimated them separately between the audited
precincts\vadjust{\goodbreak} and the un-audited ones and determined if the coefficients are statistically
different.

To see if the results are~different and to calculate the statistical significance of the
difference, it is useful to estimate the equations in the following way:
\begin{eqnarray*}
\mathrm{Votes} &=& c_0 + c_1 \quad (\mbox{vector of explanatory}
\\
&&{}\hspace*{47pt} \mbox{explaining variables})  \\
 &&{}+ c_2  \operatorname{dum}_i \quad (\mbox{vector of explanatory}\\
  &&{}\hspace*{46pt}\hspace*{16pt}\mbox{explaining
 variables}),
\end{eqnarray*}
where $c_0$, $c_1$  and $c_2$  are parameters to be
estimated and $\mathrm{dum}_i$  is a ``dummy'' variable worth 1 if we are dealing with
audited precincts and 0 if we are dealing with un-audited ones. The boxes belong to the
same random distribution if the parameters $c$ are not different from zero. The explanatory
variables utilized are the number of signatures, the number of voters registered in the REP
(Permanent Election Register) at the time of the Reafirmazo (petition re-signature
collection drive), the number of new voters registered after the Reafirmazo and the number
of voters who did not vote. We estimated the equation in logarithms.

The results are very clear, as is indicated in Table~\ref{tbl10}. The interaction term $  \mathrm{dum}_i *  \mathrm{Signatures}$ shows that the elasticity of the signatures to votes is
10.5 percent higher in the audited precincts than in the un-audited ones, that is, the
signatures collected in the audited precincts on August 18th generate 10.5 percent more YES
votes than the rest of the precincts. The statistical value of~Student's $t$ is 2.73. The
probability that this is by chance is less than~1 percent (shown~with the three asterisks
in the table). The coefficient on new voters is also different with a~level of confidence
of 1 percent whereas the coefficient with regard to the abstaining voters is different with
a level of confidence of 10 percent.

To illustrate what is unusual with this result we constructed 1,000 random samples of 200
precincts based on the universe of un-audited precincts. We estimated the same equation and
calculated the statistical value of Student's $t$ for the term $\mathrm{dum}_i  * \mathrm{Signatures}$.
The result~is shown~in Table~\ref{tbl11}. As the table shows,~a value of said statistic higher than
2.48 occurs less than 1 percent of the time. In the sample~of the audit on August 18, this
value is 2.73.

\begin{table}[t]
\caption{Do the audited precincts represent the (entire) universe of precincts?
Robust $t$ statistics in parentheses * significant at 10 \%; ** 5 \%; *** 1 \%}
\label{tbl10}
\begin{tabular*}{\tablewidth}{@{\extracolsep{4in minus 4in}}lc@{}}
\hline
 & \textbf{Log SI} \\
 \hline
$s_i$ & 0.958 \\
 & (129.46)*** \\
$\mathrm{dum}_i * s_i$ & 0.105 \\
& (2.73)*** \\
Log Electores Reafirmazo & 0.043 \\
& (4.89)*** \\
$\mathrm{dum}_i * {}$Log Electores Reafirmazo & $-$0.126 \\
& (3.06)*** \\
Log Electores Nuevos & 0.595 \\
& (23.64)*** \\
$\mathrm{dum}_i * {}$Log Electores Nuevos & 0.118 \\
& (1.30) \\
Log Electores no votantes & $-$0.459 \\
& (11.47)*** \\
$\mathrm{dum}_i * {}$Log Electores no votantes & 0.312 \\
& (1.89)* \\
$\mathrm{dum}_i * $ & 0.171 \\
& (1.51) \\
Constant & 0.254 \\
& (9.14)*** \\
Observations & 4,580 \\
$R$-squared &  0.97 \\
\hline
\end{tabular*}
\end{table}

\begin{table}[t]
\caption{Frequency distribution of the statistic of Student's $t$ value on the
parameter of signatures in 1,000 regressions estimated on the basis of 1,000 samples randomly
taken from the un-audited precincts universe}
\label{tbl11}
\begin{tabular*}{\tablewidth}{@{\extracolsep{4in minus 4in}}lcc@{}}
\hline
    &  \textbf{Percentiles} &     \textbf{Smallest}  \\
\hline
 \phantom{0}1\%  &   $-2.60853$\phantom{00}  &   $-3.342794$ \\
 \phantom{0}5\%  &  $-1.832646$\phantom{0}  &    $-3.233441$ \\
10\%  &  $-1.425525$\phantom{0}  &    $-3.053542$ \\
25\%  &  $-0.8046502$  &   $-3.053519$ \\
50\%  &  $-0.0189599$    & \\[5pt]
                 & &        \textbf{Largest} \\
\hline
75\%  &   0.7440667   &    3.232639 \\
90\%  &   1.360018\phantom{0}   &    3.658616 \\
95\%  &   1.770322\phantom{0}   &    3.975739 \\
99\%  &    2.48632\phantom{00}   &    4.010863 \\
\hline
\end{tabular*}
\begin{eqnarray*}
\mathrm{Obs}      &=&         100 \\[-2pt]
\mathrm{Sum\ of\ Wgt.}   &=&     100 \\[-2pt]
\mathrm{Mean}    &=&      -0.019166 \\[-2pt]
\mathrm{Std.\ Dev.}   &=&   1.104314 \\[-2pt]
\mathrm{Variance}   &=&    1.21950 \\[-2pt]
\mathrm{Skewness} &=&      0.074719 \\[-2pt]
\mathrm{Kurtosis}   &=&    3.04989
\end{eqnarray*}
\end{table}

We conclude that the data indicate that the audited precincts are statistically different
from the un-audited precincts. This implies that they do not form a random sample of the
entire universe of pre\-cincts (audited\vadjust{\goodbreak} and un-audited). In the audited pre\-cincts, the
signatures are transformed into a larger number of votes than in~all of~the precincts
(audited and un-audited) taken together. The probability that this occurs by coincidence is
less than~1 percent. This result tends to confirm the doubts expressed as regards the
reliability of the audit.

\subsection{Intuition}

In this section, we would like to illustrate both our theory of fraud, as well as how we
test for it.

Assume that in Florida half the precincts are Republican and half Democratic. How do we know
this? Well, first we have the results of the previous presidential election in each
precinct, which should be a good predictor of today's preferences, and we also know how
many Republicans and Democrats are registered in each precinct. Obviously, these measures
are not perfect, and they are possibly biased, but they should be related. Also assume that
on election day there are exit polls. Assume that these polls are extremely noisy and
biased.

Assume that fraud is going to be committed---in favor of the Republicans (just
an example). How can we have a perfect fraud? In the absence of an audit, electronic\vadjust{\goodbreak} fraud
is simple---when the machines connect to the central computer, the central
computer sends a program that makes the machine to report 10 percent fewer Democratic votes,
and 10 percent more Republican votes. This does not change the total number of voters but
changes the proportion.

This is undetectable, statistically speaking. The exit poll and the party registration
data will show that there is a change in votes in favor of the Republicans. The exit polls
will give a different answer, but in the end, because the exit polls are so noisy, the
blame will be given to the imperfection in the collection of the polls rather than use them
as evidence of fraud.

The only deterrent to fraud in this case is to have an audit, and the question is how we
can achieve a~kind of fraud that survives the audit. Assume that the machines leave a paper
trail of each vote and some of the machines will be audited.

Here is a fraud strategy that would be undetectable using standard statistical methods.
Assume that the election is really evenly divided between Republican and Democratic
precincts but that half of the precincts are tampered with. Let us say that 10 percent of the
votes will be shifted. The tampered result of the election will now show that
$\frac{3}{4}$ of the centers will have voted Republican and only
$\frac{1}{4}$
would have gone for the Democrats. Now, to pass the audit, the machines that will be
checked cannot belong to the set of tampered machines. This is why it is important to be
able to control the choice of the machines to be audited. Therefore, to make the fraud pass
the audit the authority draws random numbers that have $\frac{3}{4}$ weights in the
Republican precincts and $\frac{1}{4}$ in the Democratic ones. Assume this is done in the
morning of the election, so the authority knows in advance which precincts to leave
unaffected by fraud. In the end, the audit is passed, $\frac{3}{4}$ are Republican
and $\frac{1}{4}$ are Democrat.

This simple procedure---which only requires observing the results of previous
elections, or in the Venezuelan case, to observe the number of signatures and compare them
to the universe of voters---will hide fraud from an audit if the precincts are
not chosen in a truly random fashion. Here, the exit polls would give a different result,
but again, most of the discrepancy would be blamed on the exit
polls.\looseness=1

Two properties worth emphasizing are satisfied by these data. First, the mean of the
audited sample and that of the whole sample would be similar. Second, the correlation between votes
and the prior information is the same in the two samples. This, at a first glance, could look
as if this is evidence of no fraud, but that is incorrect. Remember that the correlation
between two variables is unaffected if one of the variables is multiplied by a positive
number. Hence, the correlation between the signatures and the votes in the Venezuelan case
is exactly the same as the correlation between the signatures and 90 percent of the votes.
So, a fraud of, say, 10 percent would not affect the correlation between signatures and
fraud. It would, however, affect the coefficient, which is what we check for.

Therefore, how can we detect fraud? In the audited sample, the information that existed
before---the estimates of voter preference---is a better predictor
of the actual votes than in the non-audited sample. For example, in the audited sample, if
the precinct was Democratic in the past, it has a high likelihood of being Democratic today.
But in the non-audited sample, the problem is that this relationship is weaker. In other
words, we can detect if the conditional behavior between the two samples is different, and
therefore, argue that something strange is happening in the data. This is the second test
we run.

The first test is one in which we compare the predicted error of the votes using the two
different measures of the preferences of voting. For example, the reasons why the exit poll
is an imperfect measure of voter preference are different from the reasons that make the
previous election a bad measure. For instance, one is affected by the turnout rate, while
the other one is not; one took place several months before the other one; one is collected
by the electoral committee and the other is collected by the private sector---which could
possibly have a vested interest in a particular outcome, etc. The important
aspect of our test is that the reasons why one of the measures is imperfect are different
from the reasons why the other one also is imperfect. On the other hand, if there is fraud,
then both measures have a common reason why they fail. This is our first test.

\section{Carter Center Critique}

The Carter Center issued a report entitled ``Report on an Analysis of the
Representativeness of the Second Audit Sample, and the Correlation between Petition Signers
and the Yes Vote in the Aug. 15, 2004 Presidential Recall Referendum in
Venezuela'' which is a response to our paper ``In search
of the black swan: Analysis of the Statistical Evidence of Electoral Fraud in
Venezuela.'' In preparing their response, the Carter Center never
contacted us to ask questions about our methodology or asked to see our data in order to
reproduce our results, although we offered to do so.

One important aspect of our paper is that we studied whether the sample used by the Carter
Center for the purpose of the audit was a random sample of the whole universe of automated
voting precincts. We also present what we believe to be evidence of fraud, but the Carter
Center report does not deal with this aspect of our report. They are mainly concerned with
the randomness of their subsample.

How does the Carter Center answer our claim? They make three propositions:
\begin{longlist}[(1)]
        \item[(1)] They check whether the means of the votes in the two samples are similar.
        \item[(2)] They check whether the correlation between signatures and votes is similar in the audited
and in the non-audited precincts.
        \item[(3)] They test the random number generator program used by the Electoral Council and find that
it does generate a random draw of all the precincts. They also correctly point out that the
numbers are not truly\vadjust{\goodbreak} random in the sense that the same initial seed number generates the
same sequence of numbers.
\end{longlist}

\subsection{Similar Sample Means}\label{sec5.1}

With respect to the first point, the question that the Carter Center asks is whether the
\textit{unconditional} means of the two samples are similar. By unconditional we mean that
they do not \textit{control} for the fact that precincts are different in the four
dimensions we include in our equation or in any other
dimension.\looseness=1

To see the importance of \textit{conditioning}, let us imagine that there is fraud and let
us suppose that the fraud is carried out in a large number of precincts but not in all of
them. The question is: is it possible to choose an audit sample of non-tampered centers
that has the same mean as the universe of tampered and un-tampered precincts? The answer is
obviously yes. Let us give an example using a population with a varying level of income,
say from US\$ 4,000 per year to several million. Assume that half of them have been taxed
20 percent of their income while the other half have not. Is it possible to construct an
audit sample of non-taxed individuals whose average income is similar to that of those that
have been taxed? Obviously the answer is yes. However, if one controls for the level of
education, the years of work experience and the positions they hold in the companies they
work in, it should be possible to find that the audited individuals actually had a higher
net income than the non-audited group. That is the essence of what we do.

Now, let us go back to the case in point. Precincts vary from those where the YES got more
than 90 percent of the vote to those where it got less than 10 percent. This is a very
large variation relative to the potential size of the fraud, say 10 or 20 percent. It is
perfectly feasible to choose a sample that has the same mean as the rest of the universe.

However, the non-random nature of the sample would be revealed if we compare the means but
control for the fact that each precinct is different. That is what we do and this is
the randomness test that the audited sample failed.

\subsection{Similar Correlation Coefficients}

The second check consists of comparing the correlation between signatures and votes in the
two samples, which they find to be very similar. But a simple correlation is not a test of
causality or strength.

To see this, suppose that in the audited sample there is a perfect relationship in
which\vadjust{\goodbreak}
each signature becomes 2 votes, and in the non-audited sample, because of fraud, the
relationship implies that each signature becomes only 1 vote. The correlation coefficient
in both samples is 1. This is due to the fact that the correlation coefficient is affected
by whether the two variables move up and down together, but not by whether they do so in a
relationship of 1-to-1, 2-to-1 or 10-to-1. This procedure is certainly no proof of
randomness or of the absence of fraud.

\subsection{Test of the Sample Number Generator}

The final point is that the random number generator actually generates a sample that can
potentially pick all the universe of precincts and that it was tested and appeared to
actually generate random numbers. However, there are many ways in which this kind of
analysis is weak. The most obvious one is that the program does not really generate random
numbers but a predetermined set of numbers for each seed-number that initiates the
sequence. By putting a known seed-number the Electoral Council would know beforehand which
precincts would come up, and could thus decide which precincts to leave unaltered. It is
our understanding that in the audit conducted on August 18--20, the seed number was provided
by the Electoral Council and implemented in their computer. It does not matter if, as
reported by the Carter Center, after 1,000 draws, the likelihood of any precinct being
chosen looks reasonably random. The point is that the first draw is completely
predetermined by the seed number.

\subsection{Response}

The Carter Center report does not address the two main findings of our report. It
completely disregards the evidence we put forth regarding the statistical evidence for the
existence of fraud in the statistical record. It only addresses the issues we raise
regarding the randomness of the sample used for the audit they observed on August 18--20,
2004. They show that the unconditional means between the audited sample and the rest of the
universe are similar. However, this is no proof of randomness. Conditional on the
characteristics of the precincts, we show them to be different and this result is not
challenged or addressed by the report. The report also argues that the correlation
coefficient between signatures and votes in the audited sample is similar to that in the
rest of the precincts, but this is an irrelevant statistic for this discussion. Finally,
the report checks the source code of the software used but leaves open wide avenues for
fraudulent behavior.\footnote{We do not know what happened during the audit, as we were not present.
We do know that
the sample fails the randomness test we designed. The Carter Center has nothing to say
about this fact. Paraphrasing Popper again, the Carter Center seems content in finding the
odd white swan here and there. That does not prove the proposition that the sample was
randomly chosen. We have presented a formal test of randomness and the sample fails it.
That is a black swan.}

\section{Conclusions}\label{sec6}

This report rejects certain hypotheses about fraud in the Venezuelan referendum of August
15, 2004, but not others. We did not find empirical validity for the much-discussed
hypothesis of numerical~caps. We were also unable to prove any hypothesis~that implies
differentially tampering with the (voting) machines of the same precinct. A manipulation of
this kind would alter the percentile differences in such a way as to violate the expected
variance at the precinct,~and would have been detected by this analysis.
All hypotheses of fraud must presuppose a similar tampering in all the machines of a
precinct. If this had been done in a homogeneous manner in all of the country's precincts,
none of the methods applied in this study---as well as other statistical methods---would~have been able to identify it. What allows us to develop a test of the possible
existence of fraud is precisely the heterogeneous treatment of the different precincts.
To carry out this test we used two imperfect, random and independent indicators of the
intent to vote. Our definition of fraud consists of the existence of a difference between
the voters' intended vote and the votes registered by~the CNE. Our two indicators, as
imperfect as they might be, are correlated with the intended vote of the elector, but not
with the tampering. If both are used independently in regressions to estimate the
relationship between them and the~vote count, the error term or~deviation will reflect not
only the imperfection of the instrument applied but also the fraud. If both deviations are
correlated, it~shows there~is a common element of deviation in both. This element is our
evidence of fraud. Furthermore, to be consistent~with the hypothesis of fraud, this
correlation has to be positive; that is, in those precincts where fraud was larger, both
measures would project more votes than were actually registered.
This is precisely what~we find. Our two indicators are the number of registered voters in
each precinct~who signed in the Reafirmazo\vadjust{\goodbreak} of November 2003, and the exit polls held by
S\'{u}mate and Primero Justicia on August 15th, day of the Recall Referendum. The result holds
if we control for the changes in the electoral register and the abstention rate.
Furthermore, the result holds up well to changes in the functional form of the ratio
(linear, logarithmic, percentile). The result is not due to spurious statistical effects
(errors in the variables or the possible presence of random coefficients), as they hold up
when we correct for these factors using estimators based on instrumental variables.

We note that this technique identifies fraud insofar as it is carried differentially
across precincts. It allows us to test for the presence of fraud, but it does not allow us
to estimate its magnitude as the average fraud will be reflected in the parameters of the
relationships we estimate while the differential fraud will be reflected in the error
terms. We use the error terms in the identification of fraud, not the estimated parameters.
Again, any hypothesis of fraud must presuppose that the results of all the machines in the
same precinct were tampered with proportionally. This requires some coordination mechanism. In
theory, this coordination could be in the software or in the communication with the central
computer hub. For these reasons, it is useful to point out the following precedents:
\begin{itemize}
        \item The machines had the capacity to communicate bidirectionally with the central computer
server or hub and this~communication took place.
        \item The machines communicated with the hub before printing the Certificates, which opens the
possibility that they were instructed to print results different from the real ones.
        \item The entrance of witnesses from the opposition or of the international observers to the
computer hub during election day was not allowed.
\end{itemize}
The voting system implanted in Venezuela generates voting ballots that are checked by
the~voter and placed in boxes, which are subject to audit in a random manner. A fraud
scheme must take into account how to avoid detection during an audit.

One possibility is to leave some precincts unaffected and to direct the audit to those
precincts. The choice of which precincts to affect can be done systematically or at random.
This generates two~kinds of precincts: those that were tampered with and those that were
not. Now, if the program that selects the boxes to be opened~in an audit process can be
controlled, then it will be possible to select the~boxes\vadjust{\goodbreak} of those precincts that were not
tampered with and this sample might seem random in all aspects except~as to the question of
fraud.

Our analysis~shows that the sample selected to carry out the audit on August 18, 2004 was
not random nor representative of all the precincts. In this sample, the elasticity of the
signatures compared to the votes is 10 percent higher and the possibility that this is
random is significantly less than 1 percent. We repeated our analysis randomly selecting
1,000 samples from un-audited precincts and this result does not hold.

One important fact is that the CNE refused to use the random number-generating
program~offered by the Carter Center for the August 18th audit and instead~used its own
program installed in its own computer and initialed with their own seed.

In conclusion, this study rejects certain~hypotheses of~fraud, but~indicates others that
are compatible with the statistical data.

In statistics, it is impossible to confirm a hypothesis, but it is possible to reject it.
As Karl Popper said when observing 1,000 white swans: this does not~prove the~accuracy~of
the thesis that all swans are white. Nevertheless, observing a black swan~does allow one to
reject it.

Paraphrasing Popper, our white swan represents no fraud. The results we obtain make up a
black swan. The alternate hypothesis that there was fraud is consistent with our results,
which is why we are unable to reject it.\vspace*{-3pt}

\section*{Acknowledgments}\vspace*{-3pt}

This study was requested by S\'{u}mate who also provided the databases we
used. We appreciate the great information gathering effort carried out by this
organization. We are equally indebted to a hard-working collaborator who, because of
institutional reasons, must remain anonymous. We thank Andr\'{e}s Velasco as well for his
useful comments. The opinions expressed in this report and the errors we may have incurred
are our responsibility and do not compromise either S\'{u}mate, or the universities to which we
belong.


\end{document}